\documentstyle[12pt]{article}
\begin{document}
\def\xc12pg{$^{12}C(p,\gamma)^{13}N$}
\def\c12ag{$^{12}C(\alpha,\gamma)^{16}O$}
\def\xn13pg{$^{13}N(p,\gamma)^{14}O$}
\def\be7pg{$^7Be(p,\gamma)^8B$}
\def\o14{$^{14}O$}
\def\b8{$^8B$}
\def\n16{$^{16}N$}
\thispagestyle{empty}
\begin{center}
{\normalsize Seoul National University, Muju Resort,
Korea, 20-25 August, 1993} \\
{\normalsize{\bf ASTROPHYSICS SUMMER SCHOOL(S) - 1993, LECTURES ON:}} \\
\   \\
{\Large{\bf LABORATORY MEASUREMENTS IN \\ NUCLEAR ASTROPHYSICS}} \\
\   \\
{\bf Moshe Gai}\footnote{Permanent Address: Dept. of Physics, University
of Connecticut, U-46, 2152 Hillside Dr.,Storrs, CT 06269-3046,
GAI@UConnVM.UConn.EDU, GAI@YaleVM.CIS.Yale.EDU.}
\\
{\normalsize{A.W. Wright Nuclear Structure Laboratory, Yale University,
New Haven, CT 06511}} \\
\   \\
{\bf Abstract} \\
\end{center}
After reviewing some of the basic concepts, nomenclatures and
parametrizations of Astronomy, Astrophysics and Cosmology, we introduce
a few central problems in Nuclear Astrophysics, including the hot-CNO
cycle, helium burning and solar neutrino's.  We demonstrate that
{\bf Secondary (Radioactive) Nuclear Beams} allow for considerable
progress on these problems.
\   \\
\begin{quote}
\underline{Table of Content:}
\begin{enumerate}
\item Introduction
\item Scales and Classification of Stars
\item \underline{Reaction Theory, Methods and Applications:} \\
The PP processes \\
The CNO and hot-CNO cycles \\
Burning processes in massive stars
\item \underline{Central Problems in Nuclear Astrophysics:} \\
Solar neutrino's and the \be7pg \  reaction \\
The hot-CNO cycle and the \xn13pg \ reaction \\
Helium burning and the \c12ag \ reaction
\item \underline{Solutions [with {\bf secondary (radioactive) beams}]:} \\
Direct and indirect measurements of the $^{13}N(p,\gamma)^{14}O$ \
 reaction \\
Helium burning from the beta-delayed alpha-emission of
 $^{16}N$ \\
The Coulomb Dissociation of \o14 \  and \b8 \\
\end{enumerate}
\end{quote}
\vfill\eject

\section{INTRODUCTION}

In this lecture notes I will discuss some aspects of Nuclear Astrophysics
and Laboratory measurements of nuclear processes which are of central value
for stellar evolution and models of cosmology.  These reaction rates are
important for several reason.  At first they allow us to carry out a
quantitative detailed estimate of the formation (and the origin) of the
elements; e.g. the origin of $^{11}B\ or\ ^{19}F\ $.
In these cases the understanding
of the nuclear processes involved is essential for understanding the origin
of these elements.  The understanding of the origin of these elements on
the other hand, may teach us about exotic processes such as neutrino
scattering, that may occur in stars and are believed to produce the
observed abundances of $^{11}B\ and\ ^{19}F$.
More importantly, in most cases
details of many astronomical events, such as supernova, are hidden from the
eyes of the observer (on earth).  In most cases the event is shielded by a
large mass and only telltales arrive on earth.  Such telltales include
neutrino's, or even some form of radiation.  One of the most important
telltale of an astronomical event is the abundance of the elements produced
by the thermonuclear nucleosynthesis.  And in this case it is imperative
that we completely understand the nuclear processes so that we can carry
out an accurate test of the cosmological or stellar evolution models.  In
some cases, such as in the solar model, understanding of the nuclear
processes allow for a test of the standard model of particle physics and a
search for phenomena beyond the standard model, such as neutrino masses
(magnetic moment of neutrinos) or neutrino oscillations.  Almost in all
cases one needs to understand reaction rates at energies which are
considerably below where they can be measured in the laboratory, and one
needs to develop reliable method(s) for extrapolation to low energies.

Nuclear astrophysics reached maturity mainly through the extensive work
performed at the Kellog radiation laboratory at Caltech, under the
direction of Willie Fowler and lately Charlie Barnes and Steve Koonin.
However, in spite of concentrated effort on both experimental and
theoretical sides a number of problems remain unsolved, including specific
processes in helium and hydrogen burning.  These problems are in fact
central to the field and must be addressed in order to allow for progress.
In these lectures we will address these issues and suggest new experiments
and new solutions.

{\bf Radioactive Nuclear Beams (RNB)} now available
at many laboratories around
the world have already yielded some solutions to problems of current
interest, e.g. in the Hot CNO cycle or helium burning, and appear very
promising for extending our knowledge to processes in exploding stars, such
as the rp process.  I will review in this lectures some of the current and
future applications of such secondary (radioactive) beams.

In the first section I will define some scales, classifications of stars,
nomenclatures, parameters and parametrization of relevance for nuclear
astrophysics.  We will then review some of the classical reaction chains in
burning processes and discuss traditional laboratory measurements of the
relevant nuclear reaction rates.  In the later part of the lecture series
we will develop new ideas for laboratory measurements of the required
rates, mostly carried out in the time reversed fashion.  We will
demonstrate that by measuring the reaction rates in a time reversed fashion
we construct a {\bf "Narrow Band Width Hi Fi Amplifier"}
that may allow for a
measurement of the small cross sections involved.  It is important to test
whether in fact we construct a "Hi Fidelity Amplifier", so that we are
indeed measuring rates relevant for nuclear astrophysics.  These new
techniques allow us to tackle and solve some of the oldest open questions
in Nuclear Astrophysics including the (p-wave) rate for the \c12ag
reaction of helium burning and the \be7pg reaction of importance for
the solar neutrino problem.

\section{SCALES AND CLASSIFICATION OF STARS}

Most stars have been around for long time and thus have reached a state of
statistical (hydro-dynamical) equilibrium.  Indeed most properties of stars
arise from simple hydrodynamical consideration or from the fact that stars
are nearly (but not perfect) black body radiators.  Some of the most
obviously required observational parameters of a star are its distance from
the earth and its spectrum of light emission and thus its color.

Early studies by Kepler and scientist of the Newtonian era allowed for
accurate measurements of the radii and periods of orbital motion of the
various planets, including the earth.  In these measurements the appearance
of comets were very pivotal and indeed the return of Halley's comet in
April of 1759, as reported by Harvard astronomers, was announced as a
confirmation of Newton's law of gravity.  Ironically, when Halley's comet
was late to return and did not show up between September 1758 and early
April 1759, as predicted by Edmund Halley using Newton's $1/r^2\ $ law of
gravity, Newton's law of gravity was (prematurely) declared wrong
\cite{Ho85} by
the "skeptics".  It is also worth noting that while the earliest western
record of Halley's comet is from AD 66 (that was linked to the destruction
of Jerusalem), the Chinese records go back for another 679 years, as shown
in Table 1
\cite{Wa86}.  From these measurements of radii and periods, it was
possible to determine the mass of the sun and planets with high precision;
one solar mass $M_\odot = 1.989\times 10^{30}\ $ Kg,
and $M_E = 3\mu M_\odot$.

Some of the very early measurements (developed around 1838) of the distance
of stars from the earth used the parallax method
\cite{Cl68}.  It was found that
the nearest star, Alpha-Centauri visible in the southern hemisphere (a
triple star system composed of Alpha-Centauri Proxima, A and B)
produced (after corrections for its angle) 1.52 sec of arc of angular
displacement, or a parallax of 0.76 arc sec.  Knowing the earth average orbit
radius = 149.6 MKm = 1 AU (Astronomical Unit), or approximately 8
light minutes, we calculate 1 parsec = $3.086\times 10^{16}\ $
meter, or 3.262 light
years (LY).  Indeed our closest neighbor is hopelessly far from us, at a
distance of approximately 4.2 LY.  Modern days (optical) telescopes have an
accuracy of the order of 0.01 sec of an arc and with the use of
interferometry one can improve the resolution to 0.001 sec of an arc.
Hence, the parallax method has a limited use, for stars closer then 1 kpsc.
In Fig. 1, taken from Donald Clayton's book
\cite{Cl68}, we show
characteristic distances and structures in our
galaxy.  Note that the period of rotation of our galaxy is of the order of
100 million years.

\vspace{3.5in}
\begin{center}
\underline{Table 1:}  Chinese records of Halley's Comet.
\end{center}

Early measurements performed on stars also defined its color index
\cite{Cl68},
using the response of detectors (photographic plates) with band widths
spanning the Ultraviolet, Blue and Visual spectra.  The color index is
defined as Blue magnitude minus the Visual magnitude.  Note the magnitude is
roughly proportional to -2.5 log(intensity).  Hence, hot stars are
characterized by small and in fact negative color index while cold stars
have large color index.  Astronomers are also able to correlate the color
index with the (effective) surface temperature of a star, an extensively
used parameter in stellar models.

\vspace{3.5in}
\begin{center}
\underline{Fig. 1:}  Scales of our galaxy \cite{Cl68}.
\end{center}

\subsection{Classification of Stars:}

Based on this color index one classify stars using a Hertzsprung-Russell
Diagram (after the Danish and American astronomers that developed such
diagrams around 1911-1913).  In an H-R diagram one plots the Luminosity of
a star or the bolometric magnitude (total energy emitted by a star) Vs the
surface temperature, or the color index of a star.  In Fig. 2 we show two
such H-R diagrams
\cite{Cl68,Wa92}, for star clusters with approximately equal
distance to the earth.  These stars are believed to be formed within the
same time period of approximately 100 million years, which allow for the
classification.

Stars that reside on the heavy diagonal curve are referred to as main
sequence stars.  For the main sequence stars we find the brightest star to
be with highest surface temperature and of blue color.  The main sequence
stars spend most of their life burning hydrogen and acquire mass that is
related to their luminosity: $L = const \times  M^\nu,\ with \ \nu \ =\ 3.5\
to\ 4.0$.
Stellar
evolution is most adequately described on an H-R diagram, and for example
the sun after consuming most of its hydrogen fuel will contract its core
while expanding its outer layers (to a radius that will include the earth).
The contraction at first raises the luminosity and then the sun will expand
and redden, or move up and then to the right in an H-R diagram.  At a later
stage the helium fuel will ignited in the contracted core and the sun will
move to the left on (an asymptotic branch on) the H-R diagram.  At the end
of helium burning the sun will further contract to a white dwarf, see below,
and reside
(forever) at the lower bottom left of the H-R diagram.  For main sequence
stars the luminosity is given by Planck's law $L = 4\pi R^2 \sigma T_e^4$,
(we introduced
here the effective temperature - $T_e$, since stars do not have a well defined
surface and are not perfect black body radiators).  Hence one can determine
with limited accuracy the relative radii of main sequence stars.  One
common way of measuring the radii of stars is by using the interferometry
method and the Hanbury-Brown Twiss (HBT) effect
\cite{Ba69}.  In this
measurement one measures the pair correlation function (in momentum space)
of two photons and by using boson's statistics one relates the correlation
width to the radius (of the source of incoherent photons).  For example the
sun's radius (not measured via the HBT effect) is
$R_\odot = 6.9598\times 10^8$ meters
and $R_E = 1\% R_\odot$.  While the average sun's density is
$\rho_\odot = 1.4\ g/cm^3\ (\rho_E = 5.5\ g/cm^3)$,
the central density of the sun is considerably larger, and it
was determined (from stellar hydrodynamical models) to be $\rho = 158\ g/cm^3$
with a central temperature of 15.7 MK
\cite{De64,Be68,Ba89}.  Indeed the
gravitational contraction of the sun's central core allows for the heating
of the core (from a surface temperature of approximately 6,000 K) and the
ignition of the hydrogen burning that occurs at temperatures of a few MK.
The convective zone of the sun terminates at a radius of approximately 74\%
at a temperature of approximately 2 MK and density of approximately 0.12
$g/cm^3$.

Above and to the right of the main sequence we find the {\bf Red Giant} stars
that are characterized by large luminosity and therefore they are easily
seen in the sky.  This class includes only a small number of stars, a few
percent of the known stars.  The redness of these stars arises from their
large radii and they represent a star in its later stages of evolution,
after it consumed its hydrogen fuel in the core and consist mainly of
helium.  The subgiant are believed to be stars that expand their outer
envelope while contracting their helium cores, leading to the burning of
helium.  The horizontal branch stars, on the other hand, are believed to be
at various stages of helium burning.  The supergiant stars are believed to
be stars at the advance stages of their stellar evolution and perhaps
approaching the end of their energy-generating lifetime.

In the lower left corner of the H-R diagram we find the {\bf white dwarfs}
representing approximately 10\% of known stars, which are very dense stars
of mass comparable to a solar mass, with considerably smaller radii,
comparable to the earth radius.  Due to the small surface area these stars
have large surface temperature (blue color) in order to allow them to
radiate their luminosity.  These group composes of the universe's cemetery
of stars that are inactive and simply radiate their pressure energy.  The
white dwarfs are so dense that the electron degeneracy keeps them from
collapsing
\cite{Ch84}, hence can not have a mass larger then approximately
1.4$M_\odot$, the Chandrasekhar limit,
beyond which the electron degeneracy can
not overcome the gravitational collapse.  Such massive stars (or cores of
massive stars) collapse to a neutron star or a black hole under their own
gravitational pressure.

\vspace{4in}
\begin{center}
\underline{Fig. 2:}  Hertzsprung-Russell Diagram(s) \cite{Cl68,Wa92}.
\end{center}

Cluster of stars are found very far from the sun, see Fig. 1, and they may
contain as many as $10^5-10^7$ stars in spherical distribution with a radius in
the range of approximately 10 parsec (globular cluster), other clusters
include only a few stars.   Based on the characteristics of these stars in
an H-R diagram it is believed that the age of stars in the globular cluster
is of the order of $14 \pm 3$ billion years (GY)
\cite{De91a}, or as old as the
universe itself (minus 1 GY).  Within this cluster we find a relatively
young class with blue giants as the most luminous, called population I, and
an older class with red giants as the most luminous members, called
population II.  The galactic cluster Pleiades (or Subaru in Japanese)
includes its brightest star of blue color, and the M3 globular cluster that
includes some $10^5$ stars, include its brightest star of red color.

\subsection{Age of Stars:}

First generation stars are stars that coalesced from the primordial dust
that includes approximately 24\% helium and 76\% hydrogen with traces of
lithium.  Some of these stars are small enough, and have not evolved and
are still burning hydrogen, others already converted to dwarfs.  For
example the sun (which is not a first generation star) has burned its
hydrogen fuel for the last 4.6 Billion years and will do so for
approximately 5 more Billion years.  Such first generation stars are
expected to have very small amount of elements heavier then carbon (some
times generically referred to as metals).  Thus one defines the metalicity
of a star, to be the ratio of its iron (or some time oxygen) to hydrogen
content, divided by the metalicity of the sun.  This ratio (denoted by
square brackets) is usually expressed in a log scale, typically varying
between -4 and 0.  Stars with metalicity of -3 to -4 are believed to be
primordial with ages in the range of 10 to 20 Billion years.  It should be
emphasized that while the metalicity of a star is measured on its surface,
one needs to know the core metalicity and hence one needs to introduce a
stellar atmospheric model(s), and thus these data in some cases are model
dependent.

\vspace{3.0in}
\begin{center}
\underline{Fig. 3:} Lithium abundance Vs metalicity \cite{Re88}.
\end{center}

One of the key questions in cosmology is the primordial abundance of the
elements, produced during the epoch of primordial nucleosynthesis
\cite{Sc77,Bo85}.  In Fig. 3 we show the abundance of Li Vs metalicity
\cite{Re88}.
Lithium is a very volatile element, since it readily reacts with low energy
protons via the $^7Li\ +\ p\ \rightarrow\ \alpha\ +\ \alpha$ reaction.
Consequently younger stars
show large fluctuations in Li abundance.  Fig. 3 includes stars with
metalicity as low as -3 and -3.5, and we extrapolate the Li primordial
abundance in the range of $10^{-10}\ to\ 10^{-9}$, relative to
hydrogen.  For younger
stars we expect to have an additional lithium roughly proportional to the
metalicity.  This addition arise from the fact that the inter-stellar gas,
from which younger stars coalesce, includes more produced lithium as it
exist for longer times.  The destruction of lithium in the stellar
environment would yield to a depletion in younger stars.  Indeed, the
measurements of primordial lithium abundance and D and
$^3He$ (first measured
on the moon, with the Apollo mission
\cite{Sc92}) were very pivotal for
confirming Big Bang Nucleosynthesis
\cite{Sc77,Bo85}.  In Fig. 4 we show the
predicted primordial nucleosynthesis.  In these calculations
\cite{Sc77} one
varies the ratio of photon density to baryon density to yield the observed
primordial abundances.  And with the knowledge of the photon density, from
measurements of the cosmic microwave background, one deduces the baryon
density that appears to be less then 10\% of the (critical) density required
to close the universe.  Indeed if one assumes the universe is critically
closed (as suggested in inflation models), big bang nucleosynthesis
provides some of the strongest evidence for the existence of dark matter in
the universe.

\vspace{3.2in}
\begin{center}
\underline{Fig. 4:} Big Bang Nucleosynthesis \cite{Sc77}.
\end{center}

\subsection{Distances to Far Away Stars and Galaxies:}

One of the most useful (optical) method to determine the
distances of far away stars
is with the use of Cepheid Variable stars
\cite{Cl68}.  These stars undergo
periodic variations, which are not necessarily sinusoidal.  Sir Edington
demonstrated that the pulsation of the Cepheid Variables are due to the
transfer of thermal energy of the star to mechanical energy that leads to
pulsation
\cite{Cl68}.  As a consequence the star's period of pulsation is
directly related to its mass and its luminosity.  Hence, if one measures
the apparent luminosity of a Cepheid Variable star (on earth) and its
period of pulsation one can infer the distance to the Cepheid Variable and
thus the distance of its galactic host.

One of the first uses of such an astronomical Yard Stick were carried out
by Edwin Hubble with the 100 inch telescope at Mt. Wilson observatory,
Pasadena, California, during the 1920's.  Hubble was able to identify
Cepheid Variable stars at a distance of 930,000 LY, and thus well outside
our galaxy, of diameter of approximately 100,000 LY (see Fig. 1).  Hubble
was able to show that these "Faint Nebula" correspond to galaxies different
then ours.  These nebula were catalogued by Charles Messier in 1781 (with
the Crab Nebula being M1) to allow observer to distinguish such objects
from comets.  Hubble's faint nebula are identified as the
M31 (galaxy in Andromeda) and M33
spiral galaxies.  Today the distance to the Andromeda nebula is estimated
to be over 2 MLY.

\vspace{3.5in}
\begin{center}
\underline{Fig. 5:}  Hubble's observation of $v = H\ \times\ R$.
\end{center}

Hubble later noticed that the known lines of emission from Hydrogen,
Oxygen, Calcium, etc. from stars within the same galaxy are shifted toward
the red, which he correctly interpreted as a Doppler shift.  Hubble plotted
the relative velocity (deduced from the accurate measurement of the
redshift) Vs the distance, as he could best estimate using the Cepheid
variable.  Hubble's original discovery, see Fig. 5, was of a linear
relationship between the velocities and distances $v = H\ \times\ R$,
where H is the Hubble constant.
Hubble's measurements of distance were less accurate then
possible today, and they yielded the Hubble constant
$H\ =\ 500\ Km/sec/Mpc$, as can be extracted from Fig. 5.

One of the immediate consequences of Hubble's observation was that it gave
credence to the Big Bang hypothesis, developed as one possible solution to
Einstein general relativity, in the early 20's by Alexandre Friedman in
Russia, and George Lemaitre in Belgium.  Details of Big Bang
nucleosynthesis were later worked out by Gamow in the 40's.  Incidentally
it is suggested that the name Big Bang was coined by Sir Fred Hoyle as a
way of ridiculing suggestion of George Lemaitre who referred to his own
theory as the theory of the primeval atom.  It is ironic that Hoyle who to
this date still prefers the steady state theory, got to name its rival
theory.  Unfortunately Hubble's determination of H requires a universe that
is only 2 Billion Years old.  At that time one already knew that the earth
and the solar system are much older, of the order of 4.6 Billion
years, and the Big Bang theory was rejected.  Today due to more accurate
determinations of distances (e.g. a factor of 2 change for M31, see above),
we believe that the Hubble constant is between 50 to 100 Km/sec/Mpc, with
the most probable value at 65, corresponding to a universe between 20 to 10
Billion years old with the most probable age of approximately 17 BY.

\vspace{3.5in}
\begin{center}
\underline{Fig. 6:}  Look back time VS Red Shift..
\end{center}

The expanding universe allow us to define the Fractional Red-Shift, as the
fractional stretching of wave length: $Z\ =\ \Delta \lambda / \lambda_0$,
with the Doppler shift $\omega\ =\ \omega_0 \gamma (1 + \beta cos \theta)$,
and use it to parametrize distances to far away galaxies,
radio galaxies, and quasars (young galaxies at the time of formation,
mostly composed of gas with
luminosity mostly composed of radio electromagnetic radiation).
Measurements of these far away objects allow us to look back to the instant
of the big bang as shown in Fig. 6, with the oldest known quasar at 5-10\%
of the age of the universe and the oldest radio galaxy (4C 41.17) at 10-15\%
of the age of the universe.

\subsection{The Big Bang Theory:}

The big bang theory most vividly confirmed today by the COBE satellite
mission, received one of its first strong confirmations in the work of Arno
A. Penzias and Robert W. Wilson in 1964
\cite{Pe65}, where they discovered the
isotropic emission of microwave radiation from a (cosmological) source
at a temperature of
approximately 2.7 K.  Penzias and Wilson were careful to characterize this
thermal source, but did not point to its origin from the expanding universe
of the big bang theory.  This possibility was in fact pointed out
by Peebles and Dicke.  Indeed in a
preceding paper
\cite{Di65} they demonstrated, that Penzias and Wilson measured
the expected microwave remnants of the big bang.  In fact Penzias and
Wilson who originally only designed an antenna for microwave communication
with satellites, first interpreted the continuous hum they detected from
all directions of space as arising from pigeon dropping on their antenna.

According to the big bang theory when the Universe was just
below 10 $\mu sec$,
its temperature was approximately 200 MeV and hence the universe was
composed of quarks and gluons solely.  At that time a phase transition from
the quark gluon plasma to hadron matter occurred.  At the age of
approximately 1 sec the universe had a temperature of approximately 1 MeV
(approximately 10 BK) and then the inverse beta decay process of the
neutron to the proton stopped, hence the ratio of neutrons to protons was
fixed by the temperature and the mass difference following Boltzmann law.
At approximately 100 sec after the big bang when the temperature was
approximately 100 keV the epoch of big bang nucleosynthesis commenced
\cite{Sc77,Bo85} and it lasted for a few minutes.  During big bang
nucleosynthesis as we believe today all the available neutrons were
captured to form helium, with a well understood helium fraction of
$Y_p\ =\ 24\%$.  At approximately
300,000 years when the temperature was approximately
10 eV, atoms emerged and in the same time the universe was transparent to
radiation (decoupling).  At this point the universe changed its character
from being radiation dominated to matter dominated.  As the universe
expands all characteristic dimensions expand and radiations from a source of
1 eV (10,000 K) temperature, were redshifted to larger wave lengths of
today's observed microwave radiation, corresponding to a source at 2.7 K.
Galaxies and stars we believe, first formed when the universe was
approximately 1 Billion years old.

Recent studies suggest that big bang nucleosynthesis may have in fact
occurred in an inhomogeneous inflationary universe
\cite{Ap85,Ap87,Ap88,Al87,Ma88,Bo89,Al90}.
This model predicts a low but significantly different,
abundance of heavy elements as for example produced in the rapid neutron
capture process of supernova
\cite{Ga92}.  The observation of such heavy
elements could test whether the quark-gluon to hadron phase transition is
in fact first order.  The nature of this phase transition is of great
concern for lattice QCD calculations
\cite{Br90} and indeed for understanding
QCD.  Recent observation of the abundance of $^9Be$
\cite{Ry92} and $^{11}B$, at first
appeared promising for this model but subsequent analysis showed that the
recently observed abundances (in particular the ratio $^{11}B/^9Be$) are
consistent with it arising from
spallation reaction
\cite{Wa85} and no definitive evidence
was found for these models of inhomogeneous big bang nucleosynthesis and
the standard model of big bang nucleosynthesis prevails.

\section{REACTION THEORY, METHODS AND APPLICATIONS}

The gravitational pressure in a stellar environment leads to heating of the
nuclear fuel.  When hydrogen is heated to a temperature in excess of a few
MK, it is ignited and nuclear fusion takes place.  The fusion of light
elements is the source of energy in stars and indeed the most readily
available source of energy in the universe today.  These fusion reactions
aside from "driving stars" are also the origin of the elements heavier then
helium.  The understanding of thermonuclear processes entails a complete
understanding of nuclear reactions as measured in the laboratory, as
reviewed by Willie Fowler
\cite{Fo84,Ca88} and the seminal papers of FCZ I
\cite{Fo67} and FCZ II
\cite{Fo75}.  A review of these reactions can also be found in
Rolfs and Rodney's book
\cite{Ro88}.  Usually one would like to know if a
reaction rate is sufficiently important to generate the energetic of a
stellar environment, and whether it favorably competes with other possible
reactions and decays.  In this case one needs to define the reaction time
scale, or the inverse of its rate, as we discuss below.

Consider two particles a and X, contained in a form of an ideal gas,
interacting with each other.  The reaction rate per unit volume (r) is
given by:
\begin{tabbing}
\hspace{2in} \=                         \hspace{3in} \= \\
             \> $r_{ax} = \sigma N_XJ_a$             \> (1) \\
\end{tabbing}

where $\sigma$ is the energy dependent
cross section, N is the concentration of
particles per unit volume, and J is the flux, $J_a = vN_a$, hence:
\begin{tabbing}
\hspace{2in} \=                         \hspace{3in} \= \\
             \>= $\sigma vN_XN_a $                   \> (2)
\end{tabbing}

In a star the relative velocities of a and X are distributed in a Maxwell-
Boltzmann distribution $\phi (v)$, with $ \int \phi (v)dv = 1$,
and the total thermonuclear reaction rate is given by:
\begin{tabbing}
\hspace{2in} \=                         \hspace{3in} \= \\
             \>= $N_aN_X \int v \sigma(v) \phi(v) dv$ \> \\
\   \\
             \>= $N_aN_X < \sigma v >$                \> (3)
\end{tabbing}

and for identical particle we need to introduce a further trivial
correction (to avoid double counting):
\begin{tabbing}
\hspace{2in} \=                         \hspace{3in} \= \\
             \>=$\frac{N_aNx}{(1+ \delta_{aX})} <\sigma v >$
                                                     \> (4)
\end{tabbing}
We define $\lambda = < \sigma v >$, the reaction per unit particle,
and equation 4 becomes:
\begin{tabbing}
\hspace{2in} \=                         \hspace{3in} \= \\
\hspace{1.7in}$ r_{aX}$ \>=$\lambda_{aX}\frac{N_aNx}{(1+ \delta_{aX})}$
                                                     \> (5)
\end{tabbing}

We are usually interested in characteristic time scale for the reaction and
the time that it takes to remove particle a from the stellar ensemble,
which we may want to compare for example to the beta decay lifetime of that
particle a, and we define:
\begin{tabbing}
\hspace{2in} \=                         \hspace{3in} \= \\
\hspace{1.5in}$ (\frac{\partial N_X}{\partial t})_a $
             \>=$ - \frac{N_X}{\tau_a(X)}$           \>(6a) \\
\   \\
             \>=$ - r_{aX}$
\end{tabbing}
hence:
\begin{tabbing}
\hspace{2in} \=                         \hspace{3in} \= \\
\hspace{1.5in}$ \tau_a (X)$ \>=$\frac{1}{\lambda_{aX}N_a} $\\
\   \\
                \>=$\frac{1}{< \sigma v >N_a}$        \>  (6)
\end{tabbing}

with the correct units of inverse time.  Note that the symmetry factor
$(1 +\delta_{aX}$) is now on both sides of equation 6a
and it drops out.  In order to
know if a reaction rate competes favorably with a decay rate, one needs to
evaluate equ. 6 for that reaction.  It is customary to include avogadro's
number, $N_A = 6.023 \times 10^{23}$ in equ. 6, and one
usually quotes: $N_A<\sigma v>N_a$ and $N_a$ is specified
in units of moles/Volume.

Inserting the Maxwellian into the integral in equation 6, we find:
\begin{tabbing}
\hspace{2in} \=                         \hspace{3in} \= \\
\hspace{1.3in} $<\sigma v>$
             \>= $ 4\pi (\frac{\mu}{2\pi kT})^{3/2}
                    \int v^3 \sigma(v) e^{-\frac{\mu v^2}{2kT}} \ dv$ \> (7)
\end{tabbing}
with $\mu$ the reduced mass.

Equations 6 and 7 include information from both nuclear physics (the cross
section - $\sigma$) and stellar models
(the stellar density and temperature).  The
integral is then the meeting ground for nuclear physics and stellar
physics.  Clearly the goal of nuclear astrophysics is to evaluate reactions
rates relevant to stellar environments, by use of theoretical or
experimental methods.

\subsection{The S-Factor:}

\vspace{3.5in}
\begin{center}
\underline{Fig. 7:} Cross section and S-factor for the
                $^{12}C(p,\gamma)^{13}N$ reaction \cite{Cl68}.
\end{center}

The nuclear cross section (of s-wave interacting particles) is parametrized
based on general principles of quantum mechanics, as:
\begin{tabbing}
\hspace{2in} \=                         \hspace{3in} \= \\
\hspace{1.5in} $ \sigma (E)$
             \>= $\frac{S(E)}{E} \times e^{-2\pi \eta}$ \> (8)
\end{tabbing}

where $\eta$ is the Sommerfeld parameter, $\eta = \frac{Z_1Z_2e^2}{\hbar v}$.
It is immediately clear that 1/E originates from the $\frac{\pi}{k^2}$
that appears in the expression for
the cross section in reaction theory, and the exponent accounts for the
penetration factor of the two charged particles $Z_1$ and $Z_2$.

\subsection{Non Resonant Reactions:}

\vspace{3.5in}
\begin{center}
\underline{Fig. 8:} The Gamow window predicted by equations
10 and 11 \cite{Cl68} for the \xc12pg reaction.
\end{center}

The reaction cross section and S-factor for the \xc12pg are shown in
Fig. 7.  The region of interest for stellar environment around 30 keV, (the
CNO cycle, see below) is indicated in the Figure, and it lies just beyond
the region where experiments are still possible (i.e. cross section of 20
pbarns).  It is clear that one needs to extrapolate to the energy region of
stellar conditions and the extrapolation of the S-factor allows for
additional confidence, since the S-factor varies more slowly.  Inserting
equation 8 to 7, we find:
\begin{tabbing}
\hspace{2in} \=                         \hspace{3.2in} \= \\
\hspace{1.0in} $\lambda =\ <\sigma v>$
             \>= $(\frac{8}{\mu \pi})^{1/2} \times \frac{1}{(kT)^{3/2}}
          \int S(E) \times e^{-[\frac{E}{kT} + \frac{b}{E^{1/2}}]} \ dE$\> (9)
\end{tabbing}

where we abbreviated $b = \pi Z_1Z_2 \alpha (2\mu c^2)^{1/2}$, and
$\alpha = \frac{e^2}{\hbar c}$.  And for a constant
S-factor ($S_0$) we have:
\begin{tabbing}
\hspace{2in} \=                         \hspace{3in} \= \\
             \>= $(\frac{8}{\mu \pi})^{1/2} \times \frac{S_0}{(kT)^{3/2}}
          \int e^{-[\frac{E}{kT} + \frac{b}{E^{1/2}}]} \ dE$ \> (10)
\end{tabbing}

In this case one finds that the convolution of the Maxwellian and cross
section leads to a window of most efficient energy ($E_0$) for burning, the
Gamow window, as shown in Fig. 8.

\begin{tabbing}
\hspace{2in} \=                         \hspace{3in} \= \\
\hspace{1.7in} $E_0$
             \>= $(\frac{bkT}{2})^{3/2}$ \> \\
\   \\
             \>= $1.22 ( Z_1^2 Z_2^2 \times A \times T_6^2)^{1/3} \ keV$
                                                     \> (11)
\end{tabbing}

where $T_6$ is the temperature in million degrees Kelvin,
and $A=\frac{A1A2}{(A1+A2)}$.
For example helium burning in Red Giants occurs at 200 MK ($T_6=200$), hence
the reaction \c12ag needs to be measured at energies of approximately
315 keV where helium burning is most effective.  As we shall see below this
is indeed a formidable task.

\subsection{Resonant Reactions:}

In many cases the relevant reaction rates are governed by a resonant
nuclear state.  Such states are either low lying and with narrow width, or
higher lying but acquire large width ($\Gamma \ > \ 0.1E_r$),
and can contribute
significantly to the reaction rate at low energies.  For narrow states the
contribution to the thermonuclear rate arises from the tail (at higher
temperatures) of the Boltzmann distribution and for the broad state the
thermonuclear rate arises from the tail (at lower energies) of the resonant
state.

The cross section for an interaction of particles a + b, of spins $J_1$ and
$J_2$, in a relative angular momentum state $\ell$ via an
isolated low lying (at $E_r$ close to threshold)
nuclear state, is given by the Breit-Wigner formula:

\begin{tabbing}
\hspace{2in} \=                         \hspace{3in} \= \\
\hspace{1.2in} $\sigma_{r,\ell}(a,b)$
             \>= $\frac{2 \ell + 1}{(2J_1 + 1)(2J_2 + 1)} \times
             \frac{\pi}{k^2} \times \frac{\Gamma_a \Gamma_b}{(E-E_r)^2 +
                  (\frac{\Gamma}{2})^2}$ \> (12)
\end{tabbing}

with $\Gamma_i$ the partial widths and the total width
$\Gamma = \sum_i \Gamma_i$.  The partial
widths are given by, $\Gamma_i = 2P_\ell \gamma_i^2$,
where $\gamma_i^2$ is the reduced width and $P_\ell$ the
penetrability factor, e.g. the Coulomb penetrability:

\begin{tabbing}
\hspace{2in} \=                         \hspace{3in} \= \\
\hspace{1.7in} $P_\ell$
             \>= $\frac{kR}{G_\ell^2 + F_\ell^2}$ \>
\end{tabbing}

Note that since the pentrability factor is a property of the exterior
region (of the nuclear potential), the results are independent of the
choice of the penetration factor (e.g. WKB penetration Vs. Coulomb
penetration factor), but strongly depends on the choice for nuclear radii.
One defines the statistical factor
$\omega = \frac{(2J+1)}{(2J_1+1)(2J_2+1)}$.  Note that
for most reaction rates the total width are exhausted by one particle width
(with other particle widths being energy forbidden), and the radiation
width is much smaller.  However the radiation width is the one that allows
the resonant state to de-excite to the ground state and hence form the
element of interest, as we illustrate in Fig. 9.  Cross sections of
astrophysical interest are small for energies near the resonant energy due
to the smallness of the radiation width
($\frac{\Gamma_\gamma}{\Gamma} \approx 10^{-5} - 10^{-7}$), and at
energies below resonance they are hindered by the penetrability.  It is
immediately clear that the cross section is most directly affected by the
energy of the nuclear state, the lower the resonant energy the larger the
cross section.  And the width of the state is second in this hierarchy.

For a broad state we can write the S-factor:
\begin{tabbing}
\hspace{2in} \=                         \hspace{3in} \= \\
\hspace{1.5in} $S(E)$ \>= $\frac{\pi \hbar^2}{2\mu}\ \omega\
           \frac{\Gamma_1 \Gamma_2}{(E-E_r)\ + \ \Gamma^2 /2} e^{2\pi \eta}$
                                                     \> (13)
\end{tabbing}
\begin{center}
\vspace{3.5in}
\underline{Fig. 9:} Nuclear reaction governed by a
                   (broad) nuclear state \cite{Cl68}.
\end{center}
For computational purpose it is useful to remember that $\hbar c = 197.33$
MeV fm
and $\alpha = 1/137.03$, hence $e^2 = 1.44$ MeV fm.  In
many cases the evaluation of
thermonuclear reaction rates is reduced to accurate measurements of the
partial widths that appear in equation 12
\cite{Fe89}.  When measurements are
not possible one attempts to calculate the S-factor with the use of
standard nuclear models such as sum-rules
\cite{Ga92,Al82}, and the
thermonuclear cross section could be calculated using equations 9 or 10.
We see here that the investigation of the properties of nuclear states,
i.e. Nuclear Structure Studies, are directly linked to Nuclear
Astrophysics.

For a narrow state we drive the thermonuclear rate:

\begin{tabbing}
\hspace{2in} \=                         \hspace{3in} \= \\
\hspace{1.7in} $\lambda_i$ \> = $\hbar^2 (\frac{2 \pi}{\mu kT})^{3/2}
      \omega_i \frac{\Gamma_1\Gamma_2}{\Gamma} e^{-\frac{E_r}{kT}}$
                                                     \> (14)
\end{tabbing}

And it is immediately clear that the reaction is possible due to the tail
of the Boltzmann factor, or the last term on the right hand side of
equation 14.

In the following we shall use concepts that we developed in the above
discussion of reaction theory to discuss particular processes in stars.

\subsubsection{The PP Chains:}

Stars in the main sequence like our sun, spend most of their energy
generating lifetime burning hydrogen.  The burning of hydrogen occurs in
several chains known as the PP chains
\cite{Cl68,Ba89}, as we list below:
\   \\
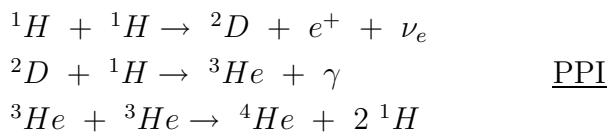
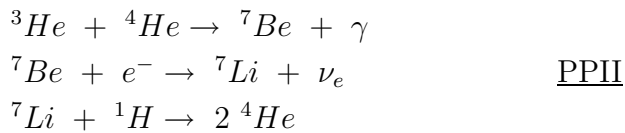
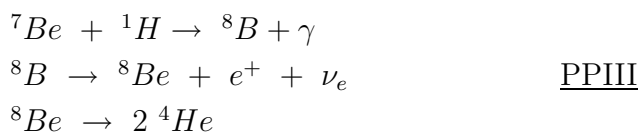
\begin{tabbing}
\hspace{1in} \= $^1H\ +\ ^1H \rightarrow\ ^2D\ +\ e^+\ +\ \nu_e$ \\
             \> $ ^2D\ +\ ^1H \rightarrow\ ^3He\ +\ \gamma$ \hspace{1in}
                                                        \underline{PPI} \\
             \> $ ^3He\ +\ ^3He \rightarrow\ ^4He\ +\ 2 \ ^1H$ \\
\   \\
\end{tabbing}
\begin{tabbing}
\hspace{2.5in} \= $^3He\ +\ ^4He \rightarrow\  ^7Be\ +\ \gamma$ \\
               \> $^7Be\ +\ e^- \rightarrow\  ^7Li\ +\ \nu_e$ \hspace{1in}
                                                         \underline{PPII} \\
               \> $^7Li\ +\ ^1H \rightarrow\ 2\ ^4He $\\
\end{tabbing}
\begin{tabbing}
\hspace{1in} \= $^7Be\ +\ ^1H \rightarrow\ ^8B + \gamma $\\
             \> $^8B\ \rightarrow\ ^8Be\ +\ e^+\ +\ \nu_e$ \hspace{1in}
                                                   \underline{PPIII} \\
             \> $^8Be\ \rightarrow\ 2\ ^4He$
\end{tabbing}

The PPI chain is the main source (98\%) of energy in the sun.  It amounts to
the fusion of 4 protons to a helium nucleus with the release of
approximately 25 MeV energy, and 95\% of the photon luminosity is produced
within 0.36 $M_\odot$ and $R\ <\ 0.21\ R_\odot$. The majority
of the energy is released in a
form of heat (kinetic energy of alpha-particles) and radiation (gamma
rays), and some energy (2.3\%) is released in the form of solar neutrino's.
The reaction rate is dictated by the weak interaction process, the first
process in the PPI chain, with a calculated S-factor $S(0)\ =\ 3.78 \pm 0.15
\times 10^{-22}$ keV-barn and linear term
coefficient $\frac{dS}{dE}\ =\ 4.2 \times 10^{-24}$ barn.
Inserting this S-factor and T = 15 MK, with the solar density of $\rho\ =\ 150
\ g/cm^3$ and $X_{He}\ =\ X_H\ =\ 0.5$, in equation 9 we
derive a reaction time, $\lambda^{-1}\ =\ 10\ BY$, i.e.
the expected lifetime of the sun.  Using available
luminosities (i.e. available beams and targets) we expect in the laboratory
at energies of astrophysical interest, an approximate rate of one p + p
interaction per year, which is clearly non measurable.  However, this rate
is considered to be reliable (within $\pm 5\%$) as it is extracted from known
weak interaction rates such as the neutron lifetime.  We also note that the
PPI neutrino luminosity (see above) is directly calculable from the total
luminosity of the sun and thus the PPI neutrino flux is considered to be
estimated with great certainty.

The burning of hydrogen release a large flux of neutrino's and with the
knowledge of the various branching ratio's and reaction rates we derive
\cite{Ba89,Ba88} for the standard solar model the neutrino flux as shown in
Fig. 10.

\vspace{3.5in}
\begin{center}
\underline{Fig. 10:}  Predicted Solar neutrino's fluxes \cite{Ba89}.
\end{center}

\subsubsection{The Solar Neutrino Problem:}

Attempts to measure solar neutrino's were carried out over the last two
decades
\cite{Ba89}.  The detection of solar neutrino's is expressed in terms of
the SNU, the Solar Neutrino Unit, which is the product of the calculated
characteristic solar neutrino flux (in units of $cm^{-2}sec^{-1}$) times the
theoretical cross section for neutrino interaction in the detector (in
units of $cm^2$).  Hence the SNU is in units of rate, events per target atom,
per second, and is chosen for convenience equal to $10^{-36}\ sec^{-1}$.
 For a
detector with $10^{31}$ atoms, one SNU yields one interaction per day.  This
counting rate is very characteristic of current solar neutrino detectors.

The first neutrino detector was constructed over two decades ago in the
Homestake mine, by Raymond Davis Jr.
\cite{Ba89} and it includes 105 gallons of
the cleaning agent carbon tetra chloride.  In this detector neutrino's with
energies above 800 keV (threshold) yield the reaction:

\begin{tabbing}
\hspace{2in} \=                         \hspace{3in} \= \\
             \> $\nu_e\ +\ ^{37}Cl\ \rightarrow\ e^-\ +\ ^{37}Ar$ \> (15)
\end{tabbing}

and the nobel gas argon is collected by bubbling helium through the tank
and collecting it in chemical adsorbers.  The decay products of the
activity of $^{37}Ar$ are counted in a
proportional counter in a low background
environment.  For this chlorine detector one predicts using Bahcall-Uhlrich
Standard Solar Model and Bahcall's various other SSM
\cite{Ba88,Ba92}
$7.9\ \pm\ 2.6\ SNU's$.  The observed rate of
the Chlorine detector over its first 7
years of operation, 1970-1977, was: $1.6\ \pm\ 0.4\ SNU$, over
the next 6 years,
1978-1984, it was: $2.2\ \pm 0.3\ SNU$, and over the last 7 years since the
detector was resuscitated with a new pump, 1986-1993, is: $2.8\ \pm\ 0.3\ SNU$.
In spite of these increases by two sigma (or more) the entire data set are
averaged over the last two decades of counting to yield the quoted rate of:
$2.2\ \pm\ 0.2\ SNU$, or $28\%\ \pm\ 3\%$ of the
rate predicted by Bahcall and Uhlrich
\cite{Ba88}.
As we discuss below other solar models that use different nuclear inputs
(see below the $S_{17}$ problem) predict a
smaller neutrino fluxes
\cite{Tu88,Tu93,Tu93a}, which are closer to the observed rate.

The Kamiokande proton decay detector (Kamiokande I) was outfitted for a
solar neutrino detector (Kamiokande II) and was used since January 1987.
It detects the Cerenkov radiation of electrons elastically scattered by the
neutrino's and it had at first a threshold of approximately 9.5, which was
later improved to 7.5 MeV.  This detector observed after approximately 1000
days of counting $46\%\ \pm\ 5\% (stat)\ \pm\ 6\% (syst)$ of
Bahcall's predicted flux
\cite{Hi90}.  Kamiokande III which consists of
improved detection systems with
larger efficiency for light collection using extensive mirrors and water
considerably cleaner with less Rn contaminant(s) and hence smaller
threshold (7 MeV), in operation since 1991
\cite{In93}, reported $56\%\ \pm\ 6\%
(stat)\ \pm\ 6\% (syst)$ of Bahcall's predicted flux.
The average of six years of
counting with the Kamioka detector amounts to $50\%\ \pm\ 4\% (stat)\ \pm\ 6%
(syst)$ of the B-U Standard Solar Model
\cite{In93} and 66\% of the SSM of Turck-Chieze
and Lopez \cite{Tu93,Tu93a}.

More recently results from gallium detectors were reported.  These
detectors have a very low threshold, of 233 keV, and hence detect the
neutrino's of the PPI chain, that extends to approximately 400 keV.  In
fact the detection of the PPI neutrino's constitute the first direct
evidence that the sun burns hydrogen as its primary source of energy.  The
(updated) SAGE collaboration reported
\cite{Ab91} $70\ \pm\ 20$ SNU's and the GALLEX
collaboration
\cite{An93} (updated) rate is: $79\pm \ 10 (stat)\ \pm 7 (syst)$,
compared to the expected rate of $132\ +20\ -17\ SNU's$.  The PPI
neutrino's contribute most of the predicted rate for Ga detectors
(approximately 55\%) and for PPI neutrino's all theoretical predictions are
within a reasonable agreement of each other, and for example Turch-Chieze
predicts $125\ \pm\ 7$ SNU expected Ga detection rate.

The Sudbury Neutrino Observatory (SNO) detector
\cite{Ew92,Mc92} is expected to
be operational in 1995.  This detector will use 1000 tons of heavy water
and is expected to have a much improved performance, as well as detect a
variety of additional neutrino processes such as neutral current interactions,
over the Kamiokande detector (or the proposed super Kamiokande detector).
This detector would also serve as a neutrino spectrometer.  A major mile
stone in its construction, the construction of the geodesic dome that will
house approximately 10,000 photo multiplier tubes, was recently achieved in
Berkeley.  We refer the reader to Art B. McDonald's article on this
important detector \cite{Mc92}.

The most popular theoretical interpretation of the hindrance of the solar
neutrino flux, by approximately a factor of 2, is the neutrino flavor
oscillation induced by a density dependent resonance effect, know as the
MSW effect
\cite{Wo78,Mi85}.  We however note that it may be difficult to
reconcile all the currently available data (from Ga, Chlorine and
Kamiokande detectors) in one theoretical frame, and the possible parameter
space is quite restricted and may require additional energy dependence of
the resonance process.

\subsubsection{The CNO cycle:}

In 1939 in a seminal paper delivered in a meeting at Washington DC, Hans
Bethe proposed that stars slightly more massive then the sun
($M\ >\ 2M_\odot$, but
with temperatures smaller then 100 MK, may generate their energy more
efficiently by burning hydrogen with the help of carbon (i.e. carbon is
acting as a catalyst), now known as the CNO cycle.  The main branch of the
CNO cycle:
\begin{center}
$^{12}C(p,\gamma)^{13}N(\beta^+)^{13}C(P,\gamma)^{14}N
(p,\gamma)^{15}O(\beta^+)^{15}N(p,\alpha)^{12}C$ \hspace{0.5in} (16) \\
$^{12}C(p,\gamma)^{13}N(p,\gamma)^{14}O(\beta^+)^{14}N
(p,\gamma)^{15}O(\beta^+)^{15}N(p,\alpha)^{12}C$ \hspace{0.5in} (16a)
\end{center}

We note that indeed in the CNO process, equation 16, like in the PP chain,
four protons were used to produce a helium nucleus, with the production of
fusion energy and the emission of electron neutrino's.  In addition the
star will now have carbon and nitrogen isotopes at various concentrations
due to this cycle.  For stars of core temperature larger then 17 MK
\cite{Be68}
the CNO cycle provides a more efficient energy source and indeed these
stars generate a large fraction of their energy through the CNO cycle as
shown in Fig. 11.

\vspace{3.5in}
\begin{center}
\underline{Fig. 11:}  The CNO - PP transition \cite{Be68}.
\end{center}

\subsubsection{The hot CNO cycle:}

The beta decay lifetime of $^{13}N$ is 863 sec and of
$^{15}O$ is 176.3 sec.  The
lifetime of $^{13}N$ is slow enough to allow for a
different branch of the CNO
cycle to develope, see equ. 16a.  Clearly if the temperatures and densities
rises, such as in explosive hydrogen stellar environments, it should be
possible to reach a point where the
\xn13pg reaction rate is fast
enough that it could favorably compete with the slow beta decay of $^{13}N$,
leading to the hot-CNO cycle, equ. 16a.  This rate is given by equation 6,
$\frac{1}{<\sigma v>N_{13}}\ <\ 863$ sec, and the
conditions are related to the reaction cross
section, density and temperatures.  One then clearly needs to know the
cross section for the reaction \xn13pg at low energies, in order to
determine the stellar conditions (density and temperature) where stars may
break into the hot CNO cycle.  This reaction is governed by the $1^-$ state at
5.17 MeV in $^{14}O$, as shown in Fig. 12.

The hot CNO cycle is found in hydrogen rich environments, at large
temperatures and densities, usually involving a binary star system(s) such
as Novae etc., hence further capture of protons and alpha-particles on
elements from the hot CNO cycle may allow for break out of the hot CNO
cycle and into the rp process
\cite{Pa89}.  In this case the production of
$^{19}Ne$ via the $^{15}O(\alpha,\gamma)^{19}Ne$ reaction,
and various related branches of the hot-CNO
cycle, play a major role.  These processes may in fact produce yet heavier
elements, such as $^{22}Ne$ and elements as heavy
as mass 60 nuclei, however we
will not cover in this lecture notes these processes.

\vspace{3.5in}
\begin{center}
\underline{Fig. 12:} Nuclear states in $^{14}O$ relevant
          to the hot-CNO cycle.
\end{center}

\subsubsection{Nucleosynthesis in Massive Stars:}

As stars consume their hydrogen fuel in the core, now composed mainly of
helium, it contracts, raising its temperature and density.  For example, in
25 solar masses stars the hydrogen burning last for 7 Million years.  At
temperatures of the order of 200 MK
\cite{Ro88}, the burning of helium sets in.
The first reaction to occur is the $\alpha\ + \alpha\ \rightarrow\ ^8Be$
due to the short
lifetime of $^8Be$ this reaction yield a
small concentration of $^8Be$ nuclei in
the star.  However, this reaction is very crucial as a stepping stone for
the next reaction that is loosely described as the three alpha-capture
process:

\begin{tabbing}
\hspace{2in} \=                         \hspace{3in} \= \\
             \> $^8Be(\alpha,\gamma)^{12}C$ \> (17)
\end{tabbing}

The formation of small concentration of $^8Be$, allows for a larger phase
space for the triple alpha-capture reaction to occur.  This reaction was
originally proposed by Fred Hoyle, as a solution for bridging the gap over
the mass 5 and 8, where no stable elements exist, and therefore the
production of heavier elements.  In fact the triple alpha capture reaction
is governed by the excited $0^+$ state in $^{12}C$ at
7.654 MeV, as shown in Fig.
13.  This state was predicted by Fred Hoyle prior to its discovery (by Fred
Hoyle and others) at the Kellog radiation lab
\cite{Fo84}.  One loosely refers
to this $0^+$ state as the reason for our existence, since without this state
the universe will have a lot less carbon and indeed a lot less heavy
elements, needed for life.  Extensive studies of properties of this state
by nuclear spectroscopist allow us to determine the triple alpha-capture
rate using equation 14.  The triple alpha process is in fact accurately
known to better then 10\%.

\vspace{3.5in}
\begin{tabbing}
\underline{Fig. 13:} \= Nuclear levels in $^{12}C$ and $^8Be$, relevant
           for the triple alpha-particle \\ \> capture reaction.
\end{tabbing}

At the same temperature range (200 MK), the produced $^{12}C$ nuclei can
undergo subsequent alpha-particle capture to form $^{16}O$:

\begin{tabbing}
\hspace{2in} \=                         \hspace{3in} \= \\
             \> $^{12}C(\alpha,\gamma)^{16}O$ \> (18)
\end{tabbing}

Unlike the triple alpha-capture reaction this reaction occurs in the
continuum, as shown in Fig. 14.  This reaction is governed by the quantum
mechanical interference of the tail of the bound $1^-$ state
at 7.12 MeV (the
ghost state) and the tail of the quasi-bound $1^-$ state at
9.63 MeV, in $^{16}O$.
As we shall see in section 4 of this lecture notes, these effects eluded
measurements of the S-factor of \c12ag reaction for the last two
decades, in spite of repeated attempts, and was only recently measured
using the time reverse process of the disintegration of $^{16}O$,
populated in
the beta-decay of $^{16}N$
\cite{Zh93,Zh93a,Zh93b,Zh93c,Bu93}.  Helium burning
lasts for approximately 500,000 years in a 25 solar mass star
\cite{Ro88}, and
occurs at temperatures of approximately 200 MK.  As we shall see below the
outcome of helium burning is very crucial for determining the final fate
of a massive star prior to its supernova collapse.

\vspace{3.5in}
\begin{center}
\underline{Fig. 14:} Nuclear levels in $^{16}O$ relevant for helium burning.
\end{center}

Stars of masses smaller then approximately 8 solar masses will complete
their energy generating life cycle at the helium burning cycle.  They will
be composed mainly of carbon and oxygen and contract to a dwarf lying
forever on the left bottom corner of the H-R diagram.  More massive stars
at the end of helium burning, commence carbon burning at a temperature of
approximately 600-900 MK.  Carbon burning lasts for 600 years in 25 solar
masses stars
\cite{Ro88}.  The main reaction process in carbon burning is the
$^{12}C(^{12}C,\alpha)^{20}Ne$ reaction,
but elements such as $^{23}Na$, and some $^{24}Mg$ are also
produced.  At temperatures of approximately 1.5 BK (or approximately 150
keV) the tail of the Boltzmann distribution allows for the photo-
disintegration of $^{20}Ne$, with an alpha-particle threshold as low as 4.73
MeV.  This reaction $^{20}Ne(\gamma,\alpha)^{16}O$ serves
as a source of alpha-particle which
are then captured on $^{20}Ne$ to form $^{24}Mg$ and $^{28}Si$.
The neon burning cycle
lasts for 1 year in a 25 solar masses stars.  These alpha-particles could
also react with $^{22}Ne$, as suggested by Icko Iben
\cite{Ib75}, to yield neutron
flux via the $^{22}Ne(\alpha,n)^{25}Mg$ reaction
and give rise to the slow capture of
neutrons and the production of the heavy elements via the s-process.  At
this point the core is rich with oxygen, and it contracts further and the
burning of oxygen commence at a temperature of 2 BK, mainly via the
reaction $^{16}O(^{16}O,\alpha)^{28}Si$, with
the additional production of elements of
sulfur and potassium.  The oxygen burning period lasts for approximately 6
months in a 25 solar masses star.  At temperatures of approximately 3 BK a
very brief (one day or so) cycle of the burning of silicon commence.  In
this burning period elements in the iron group are produced.  These
elements can not be further burned as they are the most bound (with binding
energy per nucleon of the order of 8 MeV), and they represent the ashes of
the stellar fire.  The star now resemble the onion like structure shown in
Fig. 15.

\vspace{4in}

\begin{tabbing}
\underline{Fig. 15:} \= Burning stages and onion-like structure
of a $25M_\odot$ star prior to its \\ \> supernova explosion
\cite{Ro88,We80}.
\end{tabbing}

As the inactive iron core aggregates mass it reaches the Chandrasekar limit
(close to 1.4 solar mass) and it collapses under its own gravitational
pressure, leading to the most spectacular event of a supernova.  During a
supernova the electrons are energetic enough to undergo electron capture
by the nuclei and all protons are transposed to neutrons, releasing the
gravitational binding energy (of the order of $\frac{3}{5}\frac{GM^2}{R}\
\approx\ 3\times\ 10^{53}$ ergs)
mostly in the form of neutrino's of approximately 10 MeV (and temperature
of approximately 100 GK).  As the core is now composed of compressed
nuclear matter (several times denser then nuclei), it is black to
neutrino's (i.e. absorbs the neutrino's) and a neutrino bubble is formed
for approximately 10 sec, creating an outward push of the remnants of the
star.  This outward push is believed by some to create the explosion of a
type II supernova.  During this explosion many processes occur, including
the rapid neutron capture (r process) that forms the heavier elements of
total mass of approximately $M\ \approx\ 2\%M_\odot$.

The supernova explosion ejects into the inter-stellar medium its ashes from
which at a later time "solar systems" are formed.  Indeed the death of one
star yields the birth of another.  At the center of the explosion we find a
remnant neutron star or a black hole.

One of the early records of supernova was provided by Chinese astronomers
from July 4th 1054 AD
\cite{Ro88}.  That explosion left behind a cloud known as
the Crab Nebula.  Additional observation were made by Ticho Braha and later
by his student Kepler.  These include a supernova explosion on October 8,
1604 AD in the constellation Ophiuchus, shown in Fig. 16
\cite{Ze66,Ki93} and
one in 1667 AD in the constellation Cassiopeia A.  Some speculate that the
star of Beth-Lechem may correspond to a supernova explosion that occurred
in the year 3 AD.  More recent explosions, supernova 1987A and 1993J
allowed for a more detailed examination of the nucleosynthesis as well as
the observation(s) of neutrino's from such explosions.

\vspace{3.5in}
\begin{tabbing}
\underline{Fig. 16:} \= Light curves obtained from western and
eastern historical records, \\ \> indicating a type I supernova
\cite{Ze66,Ki93}.
\end{tabbing}

It is clear from Fig. 15, that if in the process of helium burning mostly
oxygen is formed, the star will be able to take a shorter route to the
supernova explosion.  In fact if the carbon to oxygen ratio at the end of
helium burning in a 25 solar masses star, is smaller then approximately 15\%
\cite{We93}, the star will skip the
carbon and neon burning and directly proceed
to the oxygen burning.  In Fig. 17 we show the results of the neon burning
as a function of the S-factor for the \c12ag reaction
\cite{We93}, and
clearly for a cross section of the \c12ag reaction that is twice the
accepted value
\cite{Ca88} (but not 1.7 the accepted value), a 25 solar masses
star will not produce $^{20}Ne$, and the carbon burning is essentially turned
off.  This indeed will change the thermodynamics and structure of the core
of the progenitor star and in fact such an oxygen rich star is more likely
to collapse into a black hole
\cite{We93} while carbon rich progenitor stars is
more likely to leave behind a neutron star.  Hence one needs to know the
carbon to oxygen ratio at the end of helium burning (with an accuracy of
the order of 15\%) to understand the fate of a dying star and the heavy
elements it produces.

\vspace{3.5in}
\begin{tabbing}
\underline{Fig. 17:} \= Neon Formation; the turning off of carbon
burning (at twice the \cite{Ca88} \\ \> accepted value for the \c12ag
    reaction), is evident by a small \\ \> production of neon
    \cite{We93}.
\end{tabbing}

Since the triple alpha-particle capture reaction: $^8Be(\alpha,\gamma)^{12}C$
 is very well
understood, see above, one must measure the cross section of the
\c12ag reaction with high accuracy of the order of 15\% or better.
Unfortunately as we discuss in the next chapter this task was not possible
over the last two decades using conventional techniques and was only
recently tackled using radioactive beams
\cite{Zh93,Zh93a,Zh93b,Zh93c,Bu93}.

\section{CENTRAL PROBLEMS IN NUCLEAR ASTROPHYSICS}

In this chapter we shall review some of the central problems of nuclear
astrophysics.  We will review the difficulties encountered in this problems
and in some cases suggest that radioactive beams have either already
provided answers or could be used to solve these problems of the last two
decades in nuclear astrophysics.

\subsection{The $^8B$ solar Neutrino's and the \be7pg Reaction:}

The predicted PPI solar neutrino flux are less sensitive to the details of
the weak interaction nuclear process and only depends on knowledge of the
luminosity of the sun, $1.37\ kW/m^2$ at 1 AU, and $L_\odot\ =\ 3.86\
\times 10^{33}\ erg\ sec^{-1}$.
The flux of the $^8B$ solar neutrino's,
composing 75\% of those detected by Ray
Davis' chlorine detector, the Kamiokande detector and also the SNO
detector, on the other hand is very sensitive to the details of the nuclear
inputs and in particular to the \be7pg reaction, as well as the exact
solar model including opacities and central temperatures.

\vspace{3.5in}
\begin{center}
\underline{Fig. 18:} The extrapolated $S_{17}$ factor of
              Barker and Spear \cite{Ba86}.
\end{center}

The accepted value of the S-factor used by Bahcall and Uhlrich
\cite{Ba88} for
the \be7pg reaction at zero energy is, $S_{17}\ =\ 24.3$ eV-barn.  The more
recent value adopted by Bahcall and Pinsonneault
\cite{Ba92} is 22.4 eV-b.
Turck-Chieze adopted the value measured by Filippone of 20.9 eV-b
\cite{Tu88}.
This small value is one of the most significant differences between her SSM
and Bahcall's SSM.  The value of $S_{17}$ was studied in details by Barker and
Spear
\cite{Ba86} and Jonson, Kolbe, Koonin and Langanke
\cite{Jo92}.  Barker and
Spear point out to problems in the value of normalization used for the
\be7pg studies, i.e. the $^7Li(d,p)^8Li$ reaction.  They discuss the
evolution of the value of the $^7Li(d,p)^8Li$ reaction cross section measured
on the 770 keV resonance, as well as other factors and suggest the value of
$S_{17}\ =\ 17$ eV-b, or approximately a 30\% reduction in $S_{17}$
as compared to the
value adopted by Bahcall and Uhlrich, as shown in Fig. 18.  This would
imply a reduction of 30\% in the expected \b8 solar neutrino flux, indeed a
large decrease.  Johnson et al. point out to some discrepancies between
data obtained by Filippone et al.
\cite{Fi83} and the unpublished data of
Kavanagh et al. \cite{Ka69}.  Johnson et al.
\cite{Jo92} adopt the value of $S_{17}\ =\
22.4$ eV-b, as adopted by Bahcall and Pinsonneault but 8\% below the value
accepted by Bahcall and Uhlrich, as shown in Fig. 19.

\vspace{3.5in}
\begin{tabbing}
\underline{Fig. 19:} \= Comparison of the measurement of Filippone
      \cite{Fi83} and Kavanagh \cite{Ka69}. \\
                          \>  And the $S_{17}$ factor extracted
         by Johnson et al \cite{Jo92}.
\end{tabbing}

In Fig. 20 we show the ratio of the cross sections measured by Parker
\cite{Pa66}, Vaughn et al. \cite{Va70}, Kavanagh et al.
\cite{Ka69} and Filippone
et al. \cite{Fi83}.  The
data of Parker and Kavanagh et al. are in agreement with each other, as are
the data of Filippone et al. and Vaughn et al.  The two data sets are also
in good agreement on the energy dependence of the two cross sections.
However as shown in Fig. 20 the two data sets are in disagreement by
approximately 35\% on the absolute value of the cross section.

The importance of the \be7pg reaction for the evaluation of the \b8
solar neutrino flux calls for a continued interest and additional accurate
measurements of the \be7pg reaction, and in particular measurements
that can distinguish between the two absolute values of the cross sections,
see Fig. 20, are very much needed.  As John Bahcall puts it: "The
\be7pg measurement is the most important measurement in nuclear physics
today".  Whether we agree or not, it is not clear whether it is possible to
improve on the existing data, and whether in fact such measurements are
possible, but in the next chapter we discuss an interesting approach with
initial partial success, at attacking this problem with $^8B$ radioactive
beams and the use of a new technique involving the Coulomb Dissociation
(Primakoff) process.

\vspace{3.5in}
\begin{tabbing}
\underline{Fig. 20:} \= The ratio of the cross sections for \be7pg
   measured by Kavanagh \\ \> et al. \cite{Ka69} and Parker \cite{Pa66} Vs
   Filippone et al. \cite{Fi83} and Vaughn et al. [Va7o].
\end{tabbing}

\subsection{ The Hot CNO cycle and the \xn13pg Reaction:}

As we discuss in section 3.3.4, the value of the cross section of the
\xn13pg reaction at low energies is governed by the $1^-$ state at 5.17
MeV in $^{14}O$, see Fig. 12.  Hence an indirect measurement of the cross
section could be carried out by measuring its partial width.  The knowledge
of the energy of the state
\cite{De93}, its total width \cite{Ch85} and its
partial radiative width, or branching ratio for gamma decay
\cite{Fe89}, should
allow for determination of the cross section, see equations 12 and 14.
This determination turned out to be a formidable task
\cite{Wa86a,Ag89,Sm93}.
In Fig. 21 we show the radiative width extracted in these experiments
\cite{Fe89,Wa86a,Ag89,Sm93} where it is deduced from a measurement of the
branching ratio for the 5.17 MeV gamma decay and the total width of the
state
\cite{Ch85}.  Only the measurement of Fernandez et al. appears useful for
this study.  As a comment in passing we note that the use of the Energy
Weighted Dipole Sum Rule (EWDSR):

\begin{tabbing}
\hspace{2in} \=                         \hspace{3in} \= \\
\hspace{1.5in} $S_1(E1)$ \> = $\sum E(1^-)\times B(E1: 0^+\
           \rightarrow\ 1^-)$  \> \\
\   \\
             \>= $\frac{9}{4\pi}\ \frac{NZ}{A}\ \frac{e^2\hbar}{2m}$
                              \> (19)
\end{tabbing}

\vspace{3.5in}
\begin{tabbing}
\underline{Fig. 21:} \= Measured $\Gamma_\gamma(^{14}O:\ 1^-\ \rightarrow\
  0^+$) using indirect and direct methods. \\ \> Most indirect measurements,
except for the Seattle one \cite{Fe89}, \\ \> yield results less
sensitive then (even) the sum rule.  The advent of \\ \> radioactive
beams is clear.
\end{tabbing}

yield an upper limit on the radiative width of approximately 5 eV.  In this
case we assume that the $B(E1:\ 1^-\ \rightarrow\ 0^+$) does
not exhaust more then 1\% of the
EWDSR.  Note that even the largest known B(E1)'s in $^{11}Be$ and $^{13}N$
exhaust 0.09\% and 0.2\% of the EWDSR, and based on our understanding of dipole
electromagnetic decays, as first suggested by Gell-Mann and Telegdi
\cite{Ge53}
and Radicati
\cite{Ra52} for self conjugate nuclei, and with advances made by
theoretical and experimental studies of B(E1) in nuclei
\cite{Al82}, we can
estimate that the E1 decay should exhaust less then 1\% of the EWDSR, as
shown in Fig. 21.  The sum rule model then allow us to place an upper limit
on the value of the radiative width of the $1^-$ state.  In spite of a
concentrated effort and with the exclusion of the Seattle result of
Fernandez et al., it is clear that an accurate determination of the partial
widths of the $1^-$ state at 5.17 MeV in $^{14}O$ is needed.  By way of
introduction to the next chapter, we show in Fig. 21 the accurate results
obtained (in experiments that lasted for only a few days each) with
radioactive beams \cite{De91,Mo91,Kie93}.

\subsection{Helium Burning and the \c12ag Reaction:}

For understanding the process of helium burning and in particular the
oxygen to carbon ratio at the end of helium burning we must understand the
\c12ag reaction as in equation 18, at the most effective energy for
helium burning of 300 keV, see equation 11.  At this energy one may
estimate
\cite{Fo84} the cross section to be $10^{-8}$ nbarn,
clearly non measurable
in laboratory experiments.  In fact the cross section could be measured
down to approximately 1.2 MeV and one needs to extrapolate down to 300 keV,
see Fig. 22.  As we discuss below the extrapolation to low energies (300
keV) which in most other cases in nuclear astrophysics could be performed
with certain reliability, is made difficult by a few effects.

The cross section at astrophysical energies has contribution from the p and
d waves and is dominated by tails of the two bound states of $^{16}O$,
the $1^-$ at
7.12 MeV (p-wave) and the $2^+$ at 6.92 MeV (d-wave), see Fig. 14.
The p-wave
contribution arises from a detailed interference of the tail of the bound
$1^-$ state at 7.12 MeV and the broad $1^-$ state at 6.93 MeV,
see Fig. 14.  The
contribution of the bound $1^-$ state arises from its virtual alpha-particle
width, that could not be reliably measured or calculated.  Furthermore, the
tails of the quasi-bound and bound $1^-$ states interfere in the continuum and
the phase can not be determined from existing data.  Existing data could be
measured only at higher energies and therefore it does not show sensitivity
to the above questions.  Due to these reasons the cross section of the
\c12ag reaction could not be measured in a reliable way at 300 keV,
and the p-wave S-factor at 300 keV, for example, was estimated to be
between 0-500 keV barn with a compiled value of
$S_{E1}\ =\ 60\ +60\ -30$ keV-b
\cite{Ca88} and
$S_{E2}(300)\ =\ 40\ +40\ -20$ keV-b.  This
large uncertainty is contrasted by the
need to know the S-factor with 15\% accuracy, see chapter 3 and Fig.
17.  In Fig. 23 we show the results obtained over two decades for the p-
wave S-factor, with the most notable disagreement in the extracted results
of the Munster group, that quoted a very large S-factor with a small
error bar.  We refer the reader to
\cite{Zh93,Zh93a,Zh93b,Zh93c} for a
complete reference list and review of the subject.  In the next section we
will discuss new idea(s) for measuring this process (in the time reversed
fashion with $^{16}O$ disintegrating to
$\alpha\ +\ ^{12}C$) in the beta decay of $^{16}N$, or
the beta-delayed alpha-particle emission of $^{16}N$ \cite{Le93},
as performed at Yale by Zhao eta al. and by the TRIUMF collaboration.

\vspace{3.5in}
\begin{center}
\underline{Fig. 22:} The \c12ag reaction cross section \cite{Fo84}.
\end{center}

\vspace{3.5in}
\begin{center}
\underline{Fig. 23:} Measured S - factor(s) for \c12ag from
\cite{Zh93a}.
\end{center}

\section{SOLUTIONS [WITH SECONDARY (RADIOACTIVE) BEAMS]}

\vspace{3.5in}
\begin{center}
\underline{Fig. 24:} The Louvain-La-Neuve Radioactive Beam Facility.
\end{center}

\vspace{3.5in}
\begin{tabbing}
\underline{ Fig. 25:} \= The Riken-RIPS facility and the setup used for
the Coulomb Dissociation \\ \> of \b8, the
Rikkyo-Riken-Yale-Tokyo-Tsukuba-LLN collaboration \cite{Mo93}.
\end{tabbing}

In the previous chapter we have already described great advances made
with the use of radioactive beams to study the \xn13pg and the hot-CNO
cycle, see Fig. 21.  These studies were performed at the Louvain-La-Neuve
(LLN) Radioactive beam facility with $^{13}N$ radioactive beams
\cite{De91} and with $^{14}O$ radioactive beams at
Riken \cite{Mo91} and at Ganil
\cite{Kie93}.  While the facility at LLN uses an ISOL
type source and works at low energies, see Fig. 24, the facility at Riken,
see Fig. 25, as well as that at Michigan State University, see Fig. 26, use
high energy beams from fragmentation process.  In the following we shall
discuss experiments performed at the Wright Nuclear Structure at Yale
University, at MSU and at Riken.

\vspace{4in}
\begin{center}
\underline{Fig. 26:} The Michigan State University A1200 RNB facility.
\end{center}

\subsection{ The p-wave S-factor of \c12ag from the beta-delayed
       alpha-particle emission of $^{16}N$:}

The beta-delayed alpha-particle emission of $^{16}N$ allow us to study the
\c12ag reaction in its time reverse fashion, the disintegration of $^{16}O$
to $\alpha\ +\ ^{12}C$, and it provides a
high sensitivity for measuring low energy
alpha-particles and the reduced (virtual) alpha-particle width of the bound
$1^-$ state in $^{16}O$ at 7.12 MeV, see Fig. 14.  As
shown in Fig. 27, low energy
alpha-particle emitted from $^{16}N$ correspond to high energy beta's and thus
to a larger phase space and enhancement proportional to the total energy to
approximately the fifth power.  In addition the apparent larger matrix
element of the beta decay to the bound $1^-$ state provides further
sensitivity to that state.

\vspace{3.5in}
\begin{center}
\underline{Fig. 27:} Nuclear States involved in the beta-delayed
alpha-particle emission of $^{16}N$.
\end{center}

It is clear from Fig. 27 that the beta-decay in this case provides
{\bf "NARROW
BAND WIDTH HI FI AMPLIFIER"}, where the high fidelity is given by our
understanding of the predicted shape of the beta-decay's.  However, in this
case one needs to measure the beta decay, see below, with a sensitivity for
a branching ratio of the order of $10^{-9}$ or better.  In this case we have
replaced an impossible experiment (the measurement of \c12ag at low
energies) with a very hard one (the beta-delayed alpha-particle emission of
$^{16}N$).

Prediction of the shape of the spectra of delayed
alpha-particles from $^{16}N$
were first published by Baye and Descouvemont
\cite{Bay88}, see Fig. 28.  Note
the anomalous interference structure predicted to occur around 1.1 MeV, at
a branching ratio at the level of $10^{-9}$.  The
reduced alpha-particle width
of the bound $1^-$ state can be directly "read off" the spectra (if) measured
at such low energies, see Fig. 28.  We emphasize that these predictions
were published approximately five years prior to the observation of the
anomaly interference structure around 1.1 MeV
\cite{Zh93,Bu93}.  The previously
measured beta-delayed alpha-particle emission of $^{16}N$ \cite{Ne74}
was analyzed
using R-matrix theory by Barker
\cite{Ba71} and lately by Ji, Filippone, Humblet
and Koonin
\cite{Ji90}.  They conclude, as shown in Fig., 29a that the data
measured at higher energies is dominated by the quasi
bound state in $^{16}O$ at
9.63 MeV, see Fig. 14, and shows little sensitivity to the interference
with the bound $1^-$ state.  The data
measured at low energies is predicted to
have large sensitivity to the anomalous interference with the bound 1-
state, see Fig. 29.  Similar prediction were also given by a K-matrix
analysis of Humblet, Filippone, and Koonin
\cite{Hu91}, see Fig. 29b, of the
same early data on $^{16}N$ \cite{Ne74}.  We again
note that both the R-matrix
paper and K-matrix paper were published three and two years, respectively,
prior to publication of the spectra from $^{16}N$ \cite{Zh93,Bu93}.

\vspace{3.5in}
\begin{tabbing}
\underline{Fig. 28:} \= Spectrum of the beta-delayed alpha-particle
emission of $^{16}N$, predicted \\ \> by Baye and
Descouvemont \cite{Bay88}, some five years before the observation \\
                                   \> of the interference anomaly.
\end{tabbing}

As shown in Fig. 27, the beta decay can only measure the p-wave S-factor of
the \c12ag reaction, and it also includes (small) contribution from an
f-wave.  The contribution of the f-wave have to be determined empirically
and appears to be very small and leads to some (at most 15\%) uncertainty in
the quoted S-factor
\cite{Zh93,Zh93a,Zh93c}.  The extraction of the total S-
factor of the \c12ag reaction could then be performed from the
knowledge of the E2/E1 ratio which is better known then the individual
quantities.

An experimental program to study the beta-delayed alpha-particle emission
of $^{16}N$ (and other nuclei) was commenced at Yale University in early 1989.
After some four years of studies and background reduction, the first
observation of the interference anomaly was carried out in November of
1991, and presented by Zhiping Zhao in a seminar at Caltech in January
1992.  This preliminary report of the anomaly around 1.1 MeV with small
statistics (approximately 25 counts in the anomaly), has activated the
TRIUMF collaboration including Charlie Barnes of Caltech, who have
redesigned their unsuccessful search using a superconducting solenoid to
remove the beta's, and indeed in March of 1992 they also observed 9 counts in
the anomalous structure as reported by Charlie Barnes in the Ohio meeting
of the AAAS in June, 1992.  The two collaboration have then carried the
required checks and balances and analyses of the data and in November 1992,
both collaborations submitted their papers for publication in the Phys.
Rev. Lett. within ten days
\cite{Zh93,Bu93}.  While the two experiments are
very different in their production of $^{16}N$ and
detection method, and hence
acquire different systematical error(s), they quote similar S-factors, of
similar accuracy and with good agreement.  The R-Matrix analysis of the two
experiments
\cite{Zh93,Po93} yield:

\vspace{3.5in}
\begin{tabbing}
\underline{Fig. 29:} \=  (a) R-Matrix prediction [Ji91], on left and
(b) K-Matrix prediction \cite{Hu91}, \\ \> on right, of the spectrum
of beta-delayed alpha-particle emission of $^{16}N$.
\end{tabbing}

\begin{tabbing}
\hspace{2.0in} \=                         \hspace{3.0in} \= \\

\underline{Yale Result:} $\ \ \ \ \ \ \ S_{E1}(300)$ \>
 $=\ 95 \pm\ 6 (stat)\ \pm\ 28 (syst)\ keV-b$ \>  (20) \\
\underline{TRIUMF II Result:} \> $=\ 80\ +\ 18\ -26$ keV-b \>  (20a)
\end{tabbing}

In the following we will outline the Yale experiment.  The experimental
setup used at the Wright Laboratory at Yale University, is shown in Fig. 30
\cite{Zh93a,Zh93c}.  In this setup
recoiling $^{16}N$ nuclei produced with a 9 MeV
deuterium beams from the Yale-ESTU tandem and the
$^{15}N(d,p)^{16}N$ reaction,
were collected in a $Ti^{15}N$ foil (with Au backing),
tilted at $7^\circ$ with respect
to the beam, see Fig. 30.  The foil is then rotated using a stepping motor
to a counting area where the time of flight between alpha-particles
measured with an array of 9 Si surface barrier detectors, and beta-particle
measured with an array of 12 plastic scintillator, is recorded.  The
experiment is described in details
\cite{Zh93,Zh93a,Zh93c} and it arrived at a
sensitivity for the branching ratio of the
beta-decay in the range of $10^{-9}$
(to $10^{-10}$), see Fig. 31, where we show
a clean spectrum.  Other
spectra show background at the level of branching ratio of $10^{-9}$.

\vspace{3.0in}
\begin{center}
\underline{Fig. 30:} The experimental setup used in the Yale
experiment \cite{Zh93,Zh93a,Zh93c}.
\end{center}

\vspace{3.0in}
\begin{center}
\underline{Fig. 31:} Clean Time Of Flight spectrum obtained in
the Yale experiment.
\end{center}

The data measured at Yale University arise from alpha-particle that
traverse the production foil and thus need to be corrected for such
effects.  Hence, we have measured the spectra of beta-delayed
alpha-particle emission of $^{16}N$ in the absence of a foil.  This
was achieved by
implanting radioactive $^{16}N$ beams from the MSU A1200 radioactive beam
facility
\cite{Zh93b}, into a surface barrier detector and studying the
alpha-decay in the detector.  In this
experiment the absolute branching ratio for
the beta decay to the quasi bound $1^-$ state was also measured
\cite{Zh93b}.

The spectrum after correction for the foil thickness and response function
was fitted with the R-matrix formalism developed by Ji, Filippone, Humblet,
and Koonin \cite{Ji90} as shown in Fig. 32a and 32b.

An attempt to extract the total (E1 + E2) S-factor for the \c12ag
reaction from the abundance of the elements was carried out by Weaver and
Woosley
\cite{We93}, by comparing the calculated abundance of the elements to
the solar abundance.  In this calculation the \c12ag cross section is
varied between 0.5 to 3 times the value tabulated in CF88
\cite{Ca88} and listed in \cite{Ba92a}:
$S_{E1}\ =\ 60\ +60\ -30$ keV-b and $S_{E2}\ =\ 40\ +40\ -20$
keV-b.  As shown in
Fig. 33 they can reproduce the observed solar abundances only for a cross
section which is $1.7\ \pm\ 0.5$ times CF88.  If we assume
the ratio E1/E2 = 1.5,
as suggested in CF88, we derive:
\   \\
\   \\
\underline{Weaver and Woosley's Result:} $\ \ \ \ S_{E1}(300)\
          =\ 102\ \pm\ 30$ keV-b \hspace{0.5in} (21) \\
in agreement with laboratory measurements, see equ. 20.

\vspace{3.3in}
\begin{tabbing}
\underline{Fig. 32:} \= (a) R-Matrix fit \cite{Ji90} of the spectrum measured
at Yale \cite{Zh93,Zh93a,Zh93c} \\ \> (b) and range of p-wave S-factors
             accepted by the measured data.
\end{tabbing}

We conclude that a two decade problem in nuclear astrophysics in search of
the S-factor for the \c12ag reaction has perhaps (finally) seen a happy
conclusion with all three independent results in agreement.  This value can
now be used in calculation of the last stages of a progenitor (red
supergiant) star before it collapses to a supernova and in fact may allow
to derive the minimum required stellar mass for the star to leave behind a
black hole.

\vspace{3.5in}
\begin{center}
\underline{Fig. 33:} The \c12ag cross section extracted from the observed
             solar abundance of the elements \cite{We93}.
\end{center}

\subsection{The Coulomb dissociation of $^{14}O$ (hot CNO) and \b8
(solar neutrino's):}

The Coulomb Dissociation
\cite{Bau86} Primakoff \cite{Pr51} process, is the time
reverse process of the radiative capture.  In this case instead of studying
for example the fusion of a proton plus a nucleus (A-1), one studies the
disintegration of the final nucleus (A) in the Coulomb field to a proton
plus the (A-1) nucleus.  The reaction is made possible by the absorption of
a virtual photon from the field of a high Z nucleus such as $^{208}Pb$.
In this
case since $\frac{\pi}{k^2}$ for a photon is
approximately 1000 times larger than that
of a particle beam, the small cross section is enhanced.  The large virtual
photon flux (typically 100-1000 photons per collision) also gives rise to
enhancement of the cross section.  Our understanding of
the Coulomb Excitation and the virtual
photon flux allow us (as in the case of electron scattering) to deduce the
inverse nuclear process.  In this case we again construct a
{\bf "NARROW BAND
WIDTH HI FI AMPLIFIER"} to measure the exceedingly small nuclear cross
section of interest for nuclear astrophysics.  However in Coulomb
Dissociation since $\alpha Z$ approaches unity (unlike the case in electron
scattering), higher order Coulomb effects (Coulomb Post Acceleration) may
be non-negligible and they need to be understood
\cite{Ba93}.  The success of
the experiment is in fact hinging on understanding such effects and
designing the kinematical conditions so as to minimize such effects.

Hence the Coulomb Dissociation process has to be measured with great care
with kinematical conditions carefully adjusted so as to minimize nuclear
interactions (i.e. distance of closest approach considerably larger then 20
fm, hence very small forward angles scattering), and measurements must be
carried out at high enough energies (many tens of MeV/u) so as to maximize
the virtual photon flux \cite{Ja62}.  Indeed when such conditions are not
carefully selected
\cite{He91,Gaz92} the measured cross section was shown to be
dominated by nuclear effects, which can not be reliably calculated to allow
the extraction of the inverse radiative capture cross section.

Good agreement between measured cross section of radiative capture through
a nuclear state, or in the continuum, where achieved for the Coulomb
Dissociation of $^6Li$ and the
$d(\alpha,\gamma)^6Li$ capture reaction \cite{Ki91}, and the
Coulomb Dissociation of $^{14}O$ and the $p(^{13}N,\gamma)^{14}O$
capture reaction \cite{De91,Mo91,Kie93}.
In addition we note that test experiment on the Coulomb
Dissociation of $^{13}N$
\cite{Mo91} was also found to be in agreement with the
\xc12pg capture reaction.

\vspace{3.5in}
\begin{center}
\underline{Fig. 34:}  The cross section for Coulomb Dissociation
and E1 capture.
\end{center}

The Coulomb Dissociation of \b8 may provide a good opportunity for resolving
the issue of the absolute value of the cross section of the \be7pg
reaction, see chapter a and Fig. 20.  The Coulomb Dissociation yield
arise from the convolution of the inverse nuclear cross section times the
virtual photon flux.  While the first one is decreasing as one approaches
low energies, the second one is increasing (due to the small threshold of
137 keV).  Hence as can be seen in Fig. 34, over the energy region of 400
to 800 keV the predicted measured yield is roughly constant.  This is in
great contrast to the case of the nuclear cross section that is dropping
very fast at low energies, see Fig. 34.  Hence measurements at these
energies could be used to evaluate the absolute value of the cross section.

The Coulomb Dissociation process is insensitive to the M1 component of the
cross section, since the M1 virtual photon flux is smaller by approximately
$\beta^2$.  In this case the $1^+$ state in \b8 at
$E_{cm}\ =\ 632$ keV, see Fig. 18, is not
expected to be observed.  As can be seen in Fig. 35, such an M1
contribution is predicted to be approximately 10\% of the \be7pg cross
section
\cite{Ki87} measured just below 1 MeV, which yield to a 10\% correction
of the $S_{17}$-factor extracted from the
radiative capture work \cite{Ro73}, but not
the $S_{17}$ factor extracted from the Coulomb Dissociation.

\vspace{3.5in}
\begin{center}
\underline{Fig. 35:} Predicted partial wave contributions in the \be7pg
             reaction \cite{Ki87}.
\end{center}

An experiment to study the Coulomb Dissociation of \b8 was performed during
March-April, 1992, at the Riken radioactive beam facility, using the setup
shown in Fig. 25.  The experiment is a Rikkyo-Riken-Yale-Tokyo-Tsukuba-LLN
collaboration
\cite{Mo93} with Professor Tohru Motobayashi as a spokesperson,
and it will constitute the Ph.D. thesis of Mr. Iwasa, at Rikkyo University.
The radioactive beams extracted from the RIPS separator, see Fig. 25, are
shown in Fig. 36.  Indeed the results of the experiment allow us to measure
the radiative capture \be7pg cross section and preliminary results
are (only) consistent with the absolute value of the cross
section measured by Filippone
et al. \cite{Fi83} and by Vaughn et al. \cite{Va70}.  The data
are now being prepared for publication and they imply a considerable
reduction in the predicted \b8 solar neutrino flux, measured by the chlorine
and Kamiokande experiments, see Chapter 4.

\vspace{3.5in}
\begin{tabbing}
\underline{Fig. 36:} \= Radioactive beams extracted from the Riken-RIPS
facility and used in \\ \> the study of the Coulomb Dissociation of \b8,
 a Rikkyo-Riken-Yale- \\ \>Tokyo-Tsukuba-LLN collaboration \cite{Mo93}.
\end{tabbing}

We conclude that radioactive beams could be used for carefully planned
experiments to solve some of the outstanding most important problems of
nuclear astrophysics today, and hence promise a rich future for low energy
nuclear astrophysics studies.

\end{document}